\PassOptionsToPackage{dvipsnames,table}{xcolor}
\documentclass{WileyMSP-template}
\usepackage{booktabs}
\usepackage{float}
\usepackage{dblfloatfix}
\usepackage{url}
\usepackage{hyperref}
\begin{document}

\title{Secure authentication via Quantum Physical Unclonable Functions: a review}

\maketitle



\author{Pol Julià Farré$^{1}*$, Vladlen Galetsky$^{2}$, Mohamed Belhassen$^{3}$, Gregor Pieplow$^{3}$, Kumar Nilesh$^{2}$, Holger Boche$^{2}$, Tim Schr\"{o}der$^{3,4}$, Janis Nötzel$^{2}$  \& Christian Deppe$^{1}$}

\hspace{5mm}

{$^{1}$Institute of Communications Engineering, Technische Universität Braunschweig, Braunschweig, Germany\\
$^{2}$Chair of Theoretical Information Technology, Technical University of Munich, Munich, Germany\\
$^{3}$Department of Physics, Humboldt-Universit\"{a}t zu Berlin, Berlin, Germany \\
$^{4}$Ferdinand-Braun-Institut, Berlin, Germany \\
$*$Corresponding author \\

\hspace{5mm}

\textbf{Email:} $^{1}$pol.julia-farre@tu-braunschweig.de, $^{2}$vladlen.galetsky@tum.de,
$^{3}$mohamed.belhassen@physik.hu-berlin.de, $^{3}$pieplow@physik.hu-berlin.de, 
$^{2}$kumar.nilesh@tum.de,
$^{2}$boche@tum.de,
$^{4}$tim.schroeder@physik.hu-berlin.de, $^{2}$janis.noetzel@tum.de, $^{1}$christian.deppe@tu-braunschweig.de}


\dedication{}



\begin{abstract}

Quantum Physical Unclonable Functions (QPUFs) offer a physically grounded approach to secure authentication, extending the capabilities of classical PUFs. This review covers their theoretical foundations and key implementation challenges—such as quantum memories and Haar-randomness—, and distinguishes QPUFs from Quantum Readout PUFs (QR-PUFs), more experimentally accessible yet less robust against quantum-capable adversaries. A co-citation-based selection method is employed to trace the evolution of QPUF architectures, from early QR-PUFs to more recent Hybrid PUFs (HPUFs). This method further supports a discussion on the role of information-theoretic analysis in mitigating inconsistencies in QPUF responses, underscoring the deep connection between secret-key generation and authentication. Despite notable advances, achieving practical and robust QPUF-based authentication remains an open challenge.

\end{abstract}

\section{Introduction}
\label{introduction}

Physical Unclonable Functions (PUFs) are devices characterized by their inherent unclonability and their ability to produce responses that are unpredictable yet consistently reproducible for given inputs. Originally introduced in the classical context \cite{classical_puf_papu}, PUFs have since found a wide range of applications. These include secret-key generation \cite{secret_key_pufs_1, secret_key_generation_pufs_2, sharp_lower_bounds}, secure storage of cryptographic data \cite{key_storage_pufs_1, key_storage_pufs_2}, and protocols such as oblivious transfer and bit commitment \cite{review_oblivious_transfer, bit_commitment}. Central to this article, PUFs are employed in secure authentication schemes \cite{cpuf_vehicle, cpuf_applications, cpuf_for_authentication}. In the recent years, a high number of different PUF models have been found to be vulnerable against machine-learning-based attacks \cite{ml_1, ml_3, ml_4, ml_5, learning_cpufs}. As pointed out in \cite{mina_hybrid_2}, this context has motivated to bring the study of PUFs to the quantum realm, giving rise to a still young field of research with the hope and expectation of finding mathematical and robust security guarantees.

Analogous to the classical case, token-based authentication schemes can be designed relying on Quantum PUFs (QPUFs) by exploiting the fingerprint set by the unpredictable outputs (responses) of the considered QPUF when being queried with new, i.e., unqueried, inputs (challenges). However, and distinctly, QPUFs, as formally defined in \cite{Mina_unitary_qpuf}, accept quantum states as challenges, as well as they produce quantum states as responses. The first documented attempt of deriving what one might, at a glance, call a QPUF is found in \cite{readout_qpuf}. Building on this work, \cite{QRPUF_Skoric_3} also proposes a PUF-based scheme involving quantum systems. We intentionally avoid calling the mentioned proposals QPUFs because, as it will become evident in Section \ref{th_framework}, these fail to fall into the definition of a QPUF presented in \cite{Mina_unitary_qpuf}. An arguably more adequate label for the schemes proposed in the aforementioned cited works is Quantum Readout PUF (QR-PUF). As we discuss in Section \ref{review}, these enjoy a reasonable sense of practicality, as opposed to QPUFs, but require additional ad-hoc assumptions and do not seem to take full advantage of the \textit{quantumness}. That is, essentially, and as already pointed out in \cite{QPUFs_comparison} for the specific work in \cite{QRPUF_Skoric_3}, the quantum challenges and responses possible for QR-PUFs are typically mappable to classical information known by the certifier who, as opposed to that of QPUFs, becomes a trusted party.

The work in \cite{Mina_unitary_qpuf} suggests that QPUFs must be unitary or negligibly non-unitary quantum transformations, and requires the analogous of uniform randomness referred to unitary operators: Haar randomness (we refer readers unfamiliar with this concept to the text book \cite{haar_textbook} and the tutorials found in\cite{Haar_tutorial, haar_tutorial_pennylane}). Interestingly, the subsequent work \cite{Mina_Pseudorandomness} circumvents the requirement of such a costly resource for the challenge selection, while it remains needed, inconveniently, for the QPUF generation. Intimately related to that, the authors in \cite{MB_QPUF} derive two different QPUF models: the Measurement-Based QPUF (MB-QPUF) and the Ideal QPUF, constituting other solutions to circumvent the cost of requiring Haar randomness for challenge selection.

This article aims to provide the reader with a review on QPUFs, going in detail over all the aforementioned aspects within the complex development and evolution of an itself complex field of investigation. In Section \ref{defs_assums}, we revisit the theoretical framework introduced in \cite{Mina_unitary_qpuf}, and we comment on some immediate implications of it when accepting certain assumptions commonly made for PUFs. In Section \ref{req_pos}, we deliver a discussion on the current incompatibility between requirements and experimental possibilities for QPUF-based tokens, as well as the corresponding near-term perspectives. Later, in Section \ref{methodology}, we present and justify the article selection criteria used in this review, focusing on the most theoretically relevant QPUF proposals. In Section \ref{review}, we present our analysis of several articles, providing an underlying historical overview of QPUFs—illustrated as a QPUF timeline—and a detailed discussion on information-theoretic approaches used for the study and characterization of QPUFs. Section \ref{future} provides an outlook on future research and Section \ref{conclusions} concludes the article.

\section{Theoretical and practical frameworks}
\label{th_framework}

In this preliminary section, we present the key theoretical definitions and assumptions underlying the field of QPUFs, and discuss their connections to existing literature. Additionally, we examine the technical implications of this framework to clarify the practical requirements for making QPUFs operational.

\subsection{Definitions and common assumptions}
\label{defs_assums}

We begin by presenting the definition of a QPUF, originally introduced in \cite{Mina_unitary_qpuf}. It should be noted that this and the subsequent definitions are not reproduced verbatim. Rather, we introduce two deliberate adjustments. First, we allow for the possibility of incorporating more than one security parameter, thereby extending the scope of the formalism. Second, in order to keep the exposition clear and reader-oriented, we favor verbal formulations in place of fully technical mathematical expressions. These adjustments are not intended to alter the substance of the established concepts; instead, they are meant to offer a slightly different perspective that may facilitate the discussion while remaining faithful to the original framework.

\begin{definition}
\label{def_qpuf}
 $(\{\lambda_i\}, \delta_r, \delta_u, \delta_c )$-QPUF:
 
 Quantum  channel $\Lambda^{ \delta_r, \delta_u, \delta_c}_{\{\lambda_i\}\text{-}\mathrm{QPUF}}$, with $\delta_r, \delta_u, \delta_c \in [0, 1]$, and with a set of security parameters $\{ \lambda_i \mid  \lambda_i \in \mathbb{R}, \forall i\}$ that serve to adjust the desired level of security within its associated authentication  protocol. The channel $\Lambda^{ \delta_r, \delta_u, \delta_c}_{\{\lambda_i\}\text{-}\mathrm{QPUF}}$ must fulfill the following  with an  overwhelming probability:

 \begin{enumerate}
         \item $\delta_u$-uniqueness, ensuring that a generated channel $\Lambda^{ \delta_r, \delta_u, \delta_c}_{\{\lambda_i\}\text{-}\mathrm{QPUF}}$ is $\delta_u$-distinguishable (in the diamond norm \cite{diamond_norm_1, diamond_norm_2}) from any other generated instance.

        \item $\delta_r$-robustness, ensuring that $\Lambda^{ \delta_r, \delta_u, \delta_c}_{\{\lambda_i\}\text{-}\mathrm{QPUF}}$ maps $\delta_r$-indistinguishable (in fidelity \cite{Nielsen_Chuang}) challenges to $\delta_r$-indistinguishable  responses.

        \item $\delta_c$-collision resistance, ensuring that $\Lambda^{ \delta_r, \delta_u, \delta_c}_{\{\lambda_i\} \text{-}\mathrm{QPUF}}$ maps $\delta_c$-distinguishable (in fidelity) challenges to $\delta_c$-distinguishable  responses.
\end{enumerate}

\label{qpuf_def}
    
\end{definition}

\begin{remark}
    On the almost-unitaricity requirement and the QPUF-generation problem:

    In \cite{Mina_unitary_qpuf} (its Theorem 3) it is shown how the two last displayed requirements necessarily imply that  $\Lambda^{ \delta_r, \delta_u, \delta_c}_{\{\lambda_i\} \text{-}\mathrm{QPUF}}$ must  be a unitary
    or a negligibly, with respect to certain security parameters, non-unitary channel. In parallel, the uniqueness requirement mandates the existence of a  quantum-channel generating procedure available to the verifier party, and establishing a proper fingerprint. 
\label{sampling_problem}

\end{remark}

Having formally defined a QPUF, we now proceed to outline the different schemes used in the literature to leverage the QPUF potential for authentication. These schemes, ultimately aiming to achieve both the completeness and soundness properties as defined in \cite{mina_hybrid_2}, so far include (see Table \ref{token_types}):

    \begin{enumerate}
        \item The prover holds the QPUF and, in the verification phase, is asked to produce the response to a certain number of challenges (QPUF models in \cite{Mina_unitary_qpuf,client_server, t-design} or MB-QPUF in \cite{MB_QPUF}).

        \item The prover stores a response and is asked to provide it in the verification phase (Ideal QPUF in \cite{MB_QPUF}).
    \end{enumerate}

\begin{table*}[h!]
    \centering
    \caption{Comparison of the features owned by the two main types of QPUF-based authentication schemes.}
    
    \begin{tabular}{c|c|c|c}
        Protocol class & Verifier & Prover & Verification type \\ \specialrule{0.15em}{0em}{0em}
        1. QPUF as a token & Stores a set of responses & Holds the QPUF (token) & Multiple-shot verification \\ \hline 

        2. Response as a token & Holds the QPUF & Stores a response (token) & 1-shot verification\\
    
    \end{tabular}
    
    \label{token_types}
\end{table*}

In this context, the two following assumptions, inherited from the framework of classical PUFs, are typically present.

\begin{assumption}
   Unclonability: 
   
   The manufacturing process yielding $\Lambda^{ \delta_r, \delta_u, \delta_c}_{\{\lambda_i\}\text{-}\mathrm{QPUF}}$ is assumed to be uncontrollable, which prevents any adversary  from efficiently replicating it. Furthermore, the underlying physical structure of $\Lambda^{\delta_r, \delta_u, \delta_c}_{\{\lambda_i\}\text{-}\mathrm{QPUF}}$ is too complex to construct a clone of it.
\end{assumption}

\begin{assumption}

Query-based adversarial model:

It is assumed that adversaries can interact with the QPUF solely by querying it, i.e., by obtaining valid challenge-response pairs. The number of queries allowed is typically stated as a function of certain security parameters.
\end{assumption}

\begin{remark}
    On side-channel attacks:

    While the information-theoretic security of QPUFs is robustly founded on quantum principles, their physical implementations introduce significant vulnerabilities to classical side-channel attacks that exploit unintentional information leakage such as power consumption, electromagnetic radiation, or photon statistics to infer the internal physical or quantum state evolution without directly measuring the quantum challenge–response. Mitigations include implementing randomized control sequences, cryogenic or electromagnetic shielding, and secure hardware design that minimizes state-dependent leakage. However, such countermeasures often introduce decoherence or impose high resource and calibration costs, thereby complicating the balance between physical security, stability, and scalability in realistic QPUF implementations.
\end{remark}

Referring to the mentioned query-based adversarial model, the work \cite{Mina_unitary_qpuf} defines three notions of unforgeability types for QPUF-based authentication protocols.

\begin{definition}

Quantum exponential unforgeability:

Property owned by those QPUF-based authentication protocols that, under any non-previously queried challenge selection, remain unforgeable by any exponential adversary, i.e., any adversary with a number of allowed queries to the QPUF equal to an exponential function of certain security parameters.
\end{definition}

\begin{definition}

Quantum existential unforgeability:

Property owned by those QPUF-based authentication protocols that, under any non-previously queried challenge selection, remain unforgeable by any polynomial adversary. 
\end{definition}

\begin{definition}

Quantum selective/universal unforgeability:

Property owned by those QPUF-based authentication protocols that, under a constrained non-previously queried challenge selection, remain unforgeable by any polynomial adversary.
\end{definition}

\begin{remark} On the QPUF dimension and the non-inclusivity of Definition \ref{qpuf_def}:

    As already pointed out in \cite{Mina_unitary_qpuf}, we notice how the assumption made on QPUF unclonability could be violated via process tomography \cite{P_TOMOGRAPHY} within the, also assumed, query-based adversary model. In this context, exponential unforgeability is unattainable. However, an appropriate choice of security parameters, e.g. the number of qubits targeted by the QPUF channel and the number
    of responses requested during verification, can potentially lead to achieving the other two notions of unforgeability, which are more realistic and typically sufficient. Importantly, as we discuss in detail in Section \ref{review}, we thus notice how Definition 1 denies the quality of being a QPUF for the QR-PUF models \cite{readout_qpuf, QRPUF_Skoric_3, hardware_with_errors_qrpuf1, weak_QPUF, hardware_with_errors_qrpuf_2}, which target systems of fixed size thus leading to learnable quantum channels unless further assumptions are considered.
\end{remark}

\begin{remark} On the Haar-randomness problem:

\label{haar_problem}
    As stated in Remark \ref{sampling_problem}, a quantum-channel generation procedure must be available at the verifier side and, informally, it should not allow for having two \textit{too similar} instances among different runs of the QPUF-generation scheme. Now, additionally, the unclonability assumption implicitly requires such generation procedure to be non-reproducible. The security proof (Theorem 6 in \cite{Mina_unitary_qpuf}) for the weakest form of QPUF unforgeability, i.e., the selective one, motivates the cited authors to model such generation procedure as being Haar random. That is, Haar randomness paves the way toward having robust security guarantees. Nonetheless, to the best of our knowledge, there is currently no efficient method for mimicking such mathematical construction in real setups. Specifically, one can see how the two prescriptions for Haar- randomly generating unitary operators given in \cite{haar_tutorial_pennylane} introduce an exponential overhead in the number of targeted qubits.
\end{remark}

\begin{remark}
On the role of the No-cloning theorem:

    Analogous to its classical counterpart, a QPUF is not fundamentally unclonable via brute-force physical inspection. In this regard, the two assumptions stated in the current section become a source exploited in security proofs and rely on a proper QPUF engineering, which we do not discuss in this article, primarily focused on theoretic aspects. However, the No-cloning theorem for unknown quantum states \cite{no-cloning_theorem} does provide QPUFs with desirable properties not conceivable in the classical setting. Namely, within the swap-test-based \cite{Swap_test} verification stage found in \cite{Mina_unitary_qpuf}, the No-cloning theorem eliminates the need for a third trusted party, present in all classical-PUF-based authentication protocols.
\end{remark}

\subsection{QPUF requirements and today's possibilities}

\label{req_pos}

    As introduced in Section \ref{defs_assums}, a prover in a QPUF-based authentication scheme holds a QPUF or a set of quantum states that serve as responses. Such token-based authentication can be especially useful in quantum networks, where only a subset of links is authenticated \cite{chen_quantum_2021}. When a token is delivered once over an authenticated link, the token’s holder can subsequently verify their identity on unauthenticated links by sending back quantum states or the QPUF.
    
    If the tokens consist of quantum states, the distribution of these states must meet certain criteria, and the quantum memories used for storage must satisfy their own requirements. First, each state must be encodable in quantum carriers capable of transmitting information. Photons are the most common choice \cite{reiserer_cavity-based_2015, luo_recent_2023}, as they can propagate through the existing fiber-optic infrastructure. To store the photonic state, a light–to-storage interface must exist.
    
    In the absence of a one-time-authenticated channel within a network, an in-person enrollment phase is required, and quantum state storage must be portable. However, long-lived quantum memories capable of interfacing with unitary quantum operations and storing highly entangled token states \cite{volume_entangling_1,volume_entangling_2} have not yet been realized. Currently, the only viable candidates for storing simple, unentangled states are noble-gas memories, which exhibit coherence times ranging from 4 to 100 hours \cite{allmendinger_precise_2017,limes_long_2025}, but these still lack a functional interface with flying qubits.

    An extended coherence time is crucial for the majority of quantum memory-based schemes. Nevertheless, there exist scenarios in which short-lived quantum tokens are sufficient, such as in digital signatures \cite{quantum_signature} and payment transactions \cite{schiansky_demonstration_2023}, while other schemes do not require quantum memory \cite{jiang_experimental_2024,kent_practical_2022}. As summarized in Table \ref{tab:stateoftheart}, each physical platform exhibits a distinct combination of advantages and limitations. The realization of a quantum memory that meets the requirements of the QPUF-generated quantum token applications would require either a significant breakthrough within a single platform, such as the development of an engineered solid-state medium, or the implementation of a hybrid system that exploits the complementary strengths of different systems while suppressing their shortcomings.

    In the more common scenario, the prover holds the QPUF for authentication rather than storing quantum states. Quantum memories, along with interfaces between these memories and the system, are still required to store the responses. The physical implementation of random unitaries remains an active research area \cite{tang_generating_2022, kumaran_random_2024}. Research focused on designing or analyzing methods to generate specific types of randomness in quantum systems is sometimes discussed within the framework of unitary designs \cite{roy_unitary_2009}. For example, Nakata et al. \cite{nakata_efficient_2017} conjecture that a physically natural unitary design could utilize geometrically local, time-independent interactions, and propose a physical realization based on cavity Quantum Electrodynamics (QED). If the assumption of time independence is relaxed, a cavity-fed system employing random pairwise interactions on individually emitted photons could also implement a natural-design Hamiltonian \cite{thomas_efficient_2022,thomas_fusion_2024}.

    Several platforms can host such devices and distribute their states across quantum networks, provided that interactions between photons and stationary quantum systems are controllable. Candidate platforms include atoms \cite{thomas_efficient_2022,thomas_fusion_2024}, quantum dots \cite{russo_photonic_2018}, ions \cite{schupp_interface_2021}, and color centers \cite{pieplow_deterministic_2023}. Purely photonic implementations have also been proposed \cite{tang_generating_2022, zelaya_integrated_2025}.

    Any QPUF platform that does require cryogenic cooling, i.e., all of the above, except for photonic implementations, will for the time being not be portable, and may only be available at fixed nodes in a network. 
    
    Next, we provide a brief non-exhaustive overview on quantum memories and the current state of the art in photonic state storage in network applications. A more in-depth overview can be found, for example, in \cite{esguerra_rodriguez_warm_2024}.

    \subsubsection*{Working principles of quantum memories}
    This section focuses on two prevalent methods for storing quantum information via light-matter interfaces: atomic ensemble-based schemes and those that directly transduce photonic qubits into long-lived degrees of freedom within single atom-like systems. Atomic-ensemble memories typically utilize three-level atomic systems in a $\Lambda$-configuration, allowing for the coherent transfer of quantum states between photons and collective atomic spin excitations. This is achieved through mechanisms such as Electromagnetically Induced Transparency (EIT) or Raman absorption processes \cite{fleischhauer_electromagnetically_2005, shinbrough_chapter_2023}. In such setups, an incoming photonic qubit is coherently mapped onto the atomic coherence between the ground state ($\ket{g}$) and a storage state ($\ket{s}$) via interaction with a classical laser field and resonant photon absorption. The stored information is later retrieved through a coherent readout process, again driven by a control laser field \cite{lvovsky_optical_2009}.
    
    Quantum memories that directly transduce photonic qubits into single atom-like systems similarly utilize a $\Lambda$-configuration of a long-lived two-level spin system, coupled via an intermediate optically excited state. Photonic qubits can be encoded in various degrees of freedom, such as time-bin \cite{marcikic_time-bin_2002}, frequency-bin \cite{lukens_frequency-encoded_2017}, or polarization states \cite{chen_polarization_2021}, and their quantum information is mapped onto long-lived spin states through coherent photon-spin interactions \cite{reiserer_cavity-based_2015}.

    \subsubsection*{Quantum memory platforms}
     A wide variety of quantum-memory platforms exists, each platform with its own strengths and limitations. While we do  not aim to provide an exhaustive review, we offer a concise overview of the current state of the art across key platforms (see Table \ref{tab:stateoftheart}).

    \begin{enumerate}
    \item \textbf{Atomic ensembles:} Ensembles of cold or warm atoms can be employed for the storage of photonic qubits. A comprehensive overview of the current state of ensemble-based quantum memories can be found in \cite{shinbrough_chapter_2023} and \cite{esguerra_rodriguez_warm_2024}. These atomic ensembles differ primarily in their operating temperatures and coherence properties. Warm atom systems function at room temperature, eliminating the need for intricate laser cooling setups. In contrast, cold atom memories offer longer coherence times, attributed to reduced atomic collisions and narrower spectral lines.
    
    \item \textbf{Trapped Atoms:} Hyperfine states of atoms serve as reliable qubit candidates \cite{bruzewicz_trapped-ion_2019}\cite{cho_review_2015}. A key advantage of trapped atoms is their strong isolation from the environment, which significantly minimizes decoherence. Various trapping techniques are employed depending on the type of atom: ions are typically confined using oscillating radio-frequency electric fields, while neutral atoms are held using optical tweezers.
    
    \item \textbf{Color centers:}  In materials like diamond, defects in the crystal lattice, such as vacancies or the presence of foreign atoms, can give rise to color centers. These defects form energy-level structures that are optically addressable and suitable for qubit implementation \cite{orphal-kobin_coherent_2025,bhaskar_experimental_2020, ruf_quantum_2021}. Unlike trapped atoms, color centers are inherently confined within the solid-state lattice, eliminating the need for external trapping mechanisms. This simplifies the system design and allows for the integration of nanostructures around the qubit. However, the solid-state environment introduces decoherence, primarily due to interactions with other lattice defects and background noise.
    
    \item \textbf{Rare-Earth Ensemble:} In these systems, the electron spin serves as the qubit. A key advantage is that the electron resides in the atom’s inner shell \cite{shinbrough_chapter_2023}, providing enhanced shielding from environmental disturbances and thus improving coherence, which can be further extended up to one hour by incorporating the state-of the-art atomic frequency-comb technique \cite{ma_one-hour_2021}. However, this shielding also poses a challenge, as it makes the electron more difficult to manipulate, resulting in slower gate operations and more complex control requirements.
    
    \item \textbf{Fiber loops:} This approach represents a fundamentally different type of quantum memory, where the photon is stored by circulating it through a long optical fiber loop. The primary advantage is its simplicity, no quantum operations or auxiliary quantum systems are required to store the information \cite{evans_experimental_2023}. However, this method has notable limitations: photon loss accumulates over time as the photon travels through the fiber, and retrieval is not on-demand but strictly determined by the fixed length of the loop.

\end{enumerate}

\begin{table*}[t]
    \centering
    
    \caption{This table presents the state-of-the-art performance metrics for each platform. The values shown are the best reported figures and may not be directly comparable as they are not necessarily achieved under the same conditions. Superscripts denote: $*$ — $T_2^*$ coherence time measured via Ramsey interferometry, $\dagger$ — $T_2$ coherence time measured via Hahn echo or dynamical decoupling. Abbreviations: RT — Room Temperature, TW — Telecom Wavelength, NA —  Not available.}
    
    \begin{tabular}{c|c|c|c|c|c}
        Platform & Storage time & Operating temperature & Wavelength & Efficiency & Bandwidth \\ \specialrule{0.15em}{0em}{0em}
        Warm Atomic Ensemble &  1.1 $\mu$s$^*$ \cite{guo_high-performance_2019} & RT \cite{guo_high-performance_2019}& 780 nm \cite{guo_high-performance_2019}& 82\%  \cite{guo_high-performance_2019}& 170 MHz \cite{guo_high-performance_2019}\\ \hline 
        Cold atom & 4.7 ms \cite{bao_efficient_2012} & 100 $\mu$K \cite{cho_highly_2016} & 780 nm \cite{cho_highly_2016} & 87\% \cite{cho_highly_2016} & 29 Hz \cite{bao_efficient_2012} \\ \hline 
        Trapped ions & 1 hour$^*$ \cite{wang_single_2021}& RT \cite{wang_single_2021}& 369.5 nm \cite{wang_single_2021}& 98.6\% \cite{wang_single_2021}& NA \\ \hline 
        Trapped neutral atoms & 40 s$^\dagger$ \cite{barnes_assembly_2022} & 1 $\mu$K\cite{saffman_quantum_2016} & 852 nm \cite{covey_quantum_2023}& 84\% \cite{simon_interfacing_2007}& 10 kHz \cite{covey_quantum_2023}\\ \hline 
        Rare-Earth Ensemble & 1 hour$^\dagger$  \cite{ma_one-hour_2021} & 1.7 K  \cite{ma_one-hour_2021}& 580 nm \cite{ma_one-hour_2021} & 69\% \cite{hedges_efficient_2010} & 10 kHz  \cite{ma_one-hour_2021}\\ \hline 
        Color Center & 40 ms$^\dagger$ \cite{bar-gill_solid-state_2013} & RT \cite{bar-gill_solid-state_2013}& 619 nm \cite{trusheim_transform-limited_2020} & 42.3\% \cite{bhaskar_experimental_2020} & NA \\ \hline 
        
        Fiber Loops & 52 $\mu$s \cite{fook_lee_fiber_2024} & RT \cite{fook_lee_fiber_2024} & TW \cite{fook_lee_fiber_2024}& 54\% \cite{cheng_fiber-coupled_2025} & 78 kHz \cite{fook_lee_fiber_2024}
    \end{tabular}
    
    \label{tab:stateoftheart}
\end{table*}

\subsubsection*{Quantum Error Correction}
Quantum Error Correction (QEC) \cite{shor_scheme_1995, gottesman_introduction_2009} was developed to make inherently noisy quantum hardware fault-tolerant, enabling reliable quantum computation. The same principles can be extended to quantum memories to protect stored quantum information. Here, we provide a brief overview on key concepts of active QEC as applied to quantum memories, and highlight recent developments in the field. For a more detailed treatment, refer to \cite{terhal_quantum_2015, heusen_measurement-free_2024}.

\begin{figure*}
\hspace{45mm}
\includegraphics[width=0.45\linewidth]{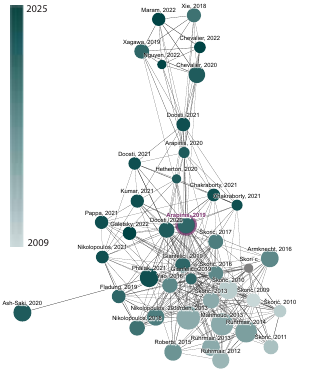}
  \caption{In examining the graph relations of \cite{Mina_unitary_qpuf} to its co-citations, we observe overlapping co-citation patterns among several related nodes.}
\label{Fig:1}
\end{figure*}

A common approach in active QEC combines multiple physical qubits (e.g., superconducting \cite{ofek_extending_2016, acharya_quantum_2025}, spin \cite{abobeih_fault-tolerant_2022}, or photonic qubits \cite{zhang_encoding_2023}) into a single logical qubit, typically within the stabilizer-code framework \cite{terhal_quantum_2015}. Below, we provide a brief overview on state-of-the-art implementations of active QEC in quantum memories.\\

\textbf{State of the art}: Google recently demonstrated a significant experimental milestone by implementing a distance-7 quantum error-correcting code using 101 qubits, achieving break-even error correction by doubling memory lifetime to $291 \pm 6~\upmu s$ compared to the longest-lived physical qubit used in the experiment \cite{acharya_quantum_2025}. In related experimental advances, the Tesseract code demonstrated distance-four encoding using just 16 physical qubits, successfully performing up to five rounds of error correction~\cite{reichardt_demonstration_2024}, while Debry \emph{et al.} encoded an error-corrected qubit in a single ion, achieving a coherence time extension factor of $1.5$\cite{debry_error_2025}. Complementing these experiments, recent theoretical research has emphasized the design of local error-correction circuits aimed at significantly extending quantum memory lifetimes\cite{park_enhancing_2025}; notably, Park \emph{et al.} proposed a low-resource code capable of preserving 12 logical qubits for nearly one million syndrome cycles using only 288 physical qubits~\cite{park_enhancing_2025}. 

Overall, the storage times and logical error rates, even with the error correction achieved to date, do not approach those of the unencoded case with conventional memories \cite{galetsky2025feasibilitylogicalbellstate}. This limitation leaves only a narrow set of specialized authentication scenarios, namely those permitting very short communication distances. Near-future applications may still arise, for example, blind quantum communication with identity authentication, in which users can authenticate only when in close proximity to the computing resource \cite{quan_verifiable_2022}.

\section{Methodology}

\label{methodology}

The procedure for identifying articles that extend QPUF research involves measuring the similarity between their reference lists, based on the assumption that articles building upon QPUFs tend to share common citations. To select the relevant articles for our study, we employed a modified version of the Jaccard similarity for sets \cite{costa2021generalizationsjaccardindex}, following these steps:

\begin{enumerate}
\item Select $m$ baseline articles $B_i$ that are known to contribute significantly to the development of QPUFs (in our case, $m = 4$).
\item Construct the reference space $\mathcal{B}$ by taking the union of the reference sets from the baseline articles:
\begin{equation}
\mathcal{B} := \bigcup_{i=1}^{m} T(B_i),
\end{equation}
where $T(B_i)$ denotes the set of references cited by the baseline article $B_i$.
\item Extract metadata and reference information from candidate articles using web Application Programming Interfaces (APIs) and/or ethical web scraping methods.
\item Compare each candidate article's reference set with the baseline reference space by matching Digital Object Identifiers (DOIs), using a cross-referencing API such as Crossref \cite{Crossref}. We define the modified Jaccard similarity as:
\begin{equation}
Sim(A_i) = \frac{|T(A_i) \cap \mathcal{B}|}{|T(A_i)|},
\end{equation}
where $T(A_i)$ is the reference set of article $A_i$.
\item An article $A_i$ is selected for further study if it satisfies the similarity threshold condition: $Sim(A_i) \geq l$, where $l$ is the predefined acceptance threshold.
\end{enumerate}

In this review, we have selected the works of QR-QPUF \cite{experimental_qrpuf}, QPUF \cite{Mina_unitary_qpuf,MB_QPUF}, and Hybrid PUFs (HPUFs) \cite{mina_hybrid_2} as our baseline articles, as they introduce novel theoretical perspectives related to the QPUF topic. It is important to note a selection bias in our choice of baselines, as we primarily focus on approaches that emphasize theoretical protocol innovation. Furthermore, for our analysis, we have set the acceptance threshold value to $l = 0.1$.
The outcome of the article selection is presented in Table \,\ref{tab:quantum_puf} within the Appendix, where baseline articles are highlighted in yellow, and accepted articles for review  are shaded in light gray. To better illustrate the relationships between articles, Figure \,\ref{Fig:1} shows a co-citation network, with the article \cite{Mina_unitary_qpuf} serving as a baseline reference.

\newpage

\section{Review}
\label{review}

\begin{figure*}[h!]
\centering
\includegraphics[width=1.0\textwidth,clip]{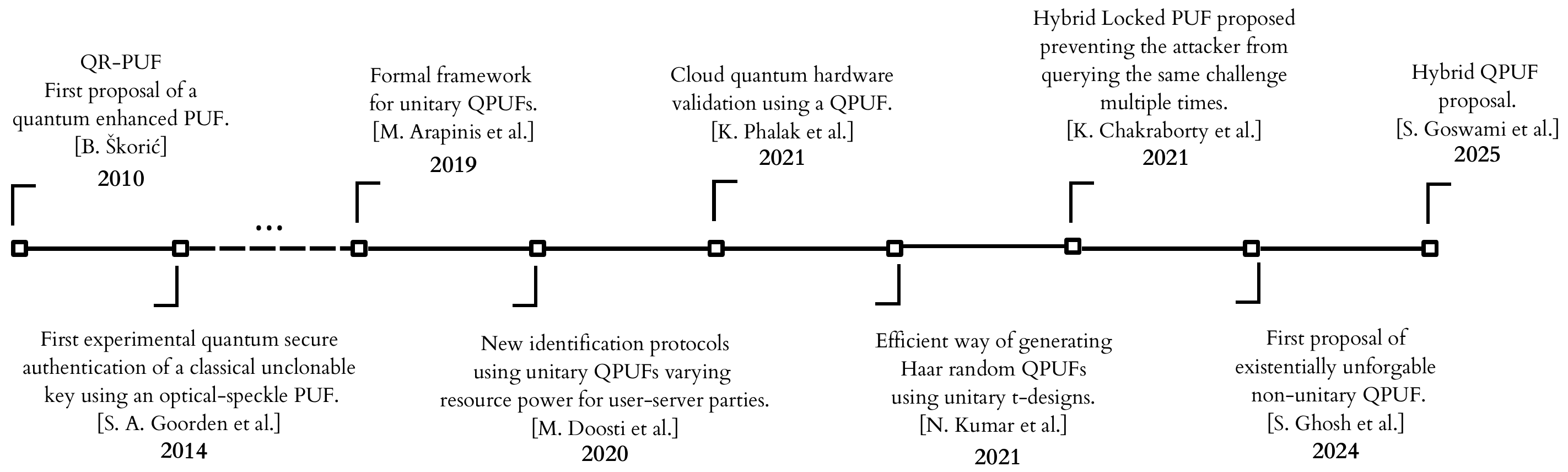}

\caption{Timeline with some of the major theoretical and hardware developments in the field of QPUFs. }
\label{Fig:Timeline}    
\end{figure*}

In this section, we review the development of Quantum Physical Unclonable Functions (QPUFs), with a timeline of key milestones illustrated in Figure \ref{Fig:Timeline}. We begin with Quantum-Readout PUFs (QR-PUFs), which are foundational to the field and noteworthy for their relatively modest hardware requirements compared to QPUFs. We introduce the initial QR-PUF proposals and highlight the first significant experimental demonstration. We also identify recurring features across various implementations, such as reliance on a trusted third party. With regard to QPUFs, our focus is on the body of work that builds upon the framework introduced in \cite{Mina_unitary_qpuf}, examining the key enhancements and new directions proposed. A central point of discussion is whether these follow-up models satisfy the criteria established in Definition \ref{def_qpuf}. In instances where they do not, we analyze the implications for the associated security properties. Moreover, we introduce another class of PUF models—known as Hybrid PUFs (HPUFs)—which offer an alternative approach by incorporating classical PUF architectures at their core. Finally, one of the articles selected for review ultimately inspired a comprehensive overview of QPUF analyses based on information-theoretic approaches and tools.

\subsection{QR-PUFs}

\subsubsection*{Birth and development of QR-PUFs}

The first QR-PUF was coined in \cite{readout_qpuf}. In this work, the author assumes the existence of a physically unclonable device capable of performing quantum unitary evolutions. Two further assumptions are, first, that distinct generated unitaries are unique and distinguishable, and second, that attackers cannot emulate such unitary evolution with a sufficiently small time delay. Under these  assumptions, two protocols are derived for remote authentication, each assuming different measurement capabilities at the verifier side. Moreover, another protocol achieving both unidirectional and mutual authentication for a Quantum Key Distribution (QKD) scheme is proposed. 

The author of \cite{readout_qpuf} highlights, as one of their main strengths, that the presented QR-PUF protocols do not rely on trusted remote readers. This makes them secure against emulation attacks, but, however, it comes at the price of the strong assumptions made. Specifically, and as discussed in the cited article itself, assuming the  impossibility of efficiently building a quantum emulator of the  QR-PUF plays a paramount role and it is unclear whether it is a plausible premise. Intimately related to that, the work in \cite{clone_attacks_qrpuf} presents a cloning-based attack, further outlining  the importance of the critical assumptions present within the discussed model.

Readers interested in security analyses concerning the first QR-PUF prototypes are referred to \cite{security_challenge, quadrature_attacks}. The former study examines the security of the models introduced in \cite{readout_qpuf}, specifically against challenge estimation attacks. The latter presents a notable experimental implementation of a QR-PUF in a practical setup \cite{experimental_qrpuf}. In particular, it evaluates the performance of a QR-PUF authentication protocol in an optical setting, accounting for quadrature-based attack strategies. The authors introduce quantitative metrics to effectively distinguish between legitimate clients and potential attackers.

Finally, we note that in an extension of the original model proposed in \cite{readout_qpuf}, the work in \cite{QRPUF_Skoric_3} presents a scheme that leverages a QR-PUF to authenticate the transmission of both classical and quantum information.

\subsubsection*{The QR-PUF theoretical framework}

Within the formal aspect of QR-PUFs, the  authors of \cite{framework_readout} propose a theoretical framework in order to establish common quantifiable  notions of security for distinct instances of either classical PUFs or QR-PUFs. We notice  that under the considerations of the cited work, the set of challenges and responses for a QR-PUF can be fully characterized by classical information owned by the certifier, who becomes a trusted party. This trait, as pointed out in \cite{QPUFs_comparison} for the already introduced work in \cite{QRPUF_Skoric_3}, seems to encompass all QR-PUF models, and sets them apart from QPUFs. Moreover, \cite{framework_readout} establishes the notions of robustness and unclonability, which resemble, respectively, the security notions of completeness and soundness \cite{mina_hybrid_2}. The authors further distinguish between two kinds of unclonability. The first kind refers to  physical unclonability,  while the second kind, under the label of mathematical unclonability, formalizes the feature of non-learnability of a QR-PUF via query-based attacks. Importantly, quantifying the latter requires no reference to quantum properties, since it is done with respect to the classical characterization of the QR-PUF.

\subsection{QPUFs}

\label{review_qr-pufs}

\subsubsection*{Polysemy of the term QPUF in the literature}

The introduced theoretical framework on QR-PUFs, together with the one presented for QPUFs in Section \ref{defs_assums}, raises the question of whether the models presented in \cite{hardware_with_errors_qrpuf1, weak_QPUF} should be regarded as QR-PUF proposals, even though they are presented as QPUFs. That is, while the challenge-response mapping of the PUFs presented in the  two cited articles stems from a quantum state preparation and measurement, such mapping is known by the certifier, and  encoded as classical information. Both works propose a model that  defines 1-qubit random rotations as challenges and monitors the output measurement histograms, which become the responses. Notably, these proposed schemes introduce the concept of using variational circuits constrained to a specific architecture, capable of handling an arbitrary number of input qubits. Such number  loosely functions as a security parameter, because the circuit complexity, and thus the learnability overhead, only scales polynomially with the number of input qubits.

Also labeled as a QPUF, the model proposed  by the authors in \cite{hardware_with_errors_qrpuf_2} introduces the idea of associating a unique fingerprint to quantum hardware devices due to the uncontrollable variations in their qubit frequency, defined by its excited-to-ground-state energy difference. Remarkably, the authors display a desirable hamming weight within the keys generated via their fuzzy extraction \cite{fuzzy_extractor} scheme, as well as a desirable hamming distance exhibited by different generated keys. Arguably, this contribution can also be seen as a QR-PUF model because the proposed set of  challenges and responses is mappable to classical information owned by the certifier.

\subsubsection*{QPUF schemes building on \cite{Mina_unitary_qpuf}}

With regard to QPUF models that adjust to the definitions provided in \cite{Mina_unitary_qpuf}, we find three relevant works (see Table \ref{tab:improvements/drawbacks} for a schematic comparative):

\begin{enumerate}
    \item Firstly, the article \cite{client_server}  proposes using a QPUF device for client-server authentication. This work includes two authentication protocols with distinct hardware requirements: one being suitable for server authentication, with a Low Resources Verifier (LRV), and the other being suitable for client authentication, with  a High Resources Verifier (HRV). The former protocol introduces major variations to the initial scheme found in \cite{Mina_unitary_qpuf}, owning extra steps that include QPUF-device exchange and quantum state shuffling. At the verifier side, no quantum measurements are needed to be carried out, but quantum memories are still required. As for the latter defined protocol, the main novelty introduced is that it allows one to choose between two different testing algorithms, one relying on the ordinary swap test, as originally conceived, and the other one relying on the so-called generalized swap (gswap) test. 

In the two mentioned types of test, exponential security is achieved in the following sense: the probability of having a successful forgery for a polynomial adversary, in the number of targeted qubits, decreases exponentially with the security parameter $N$, i.e., the number of different challenges tested per round. Nevertheless, it is worth stressing that the completeness property, i.e., the assurance that legitimate provers are accepted, is highly dependent on the quantum noiseless assumption. That is, for the swap-test-based verification algorithm, a tiny amount of quantum noise brings the probability of true acceptance to an exponentially, in the number of responses $M$ tested per each different challenge, low value. If  gswap test is chosen instead, such problem is mitigated, but remains a concern to be addressed in real scenarios.

More specifically, let us fix $N=1$ challenges per verfication-decision round, and let us assume that the fidelities $\{F_i\}_{i=1}^M$ between the responses to the fixed challenge generated by a legitimate prover and those stored by the verifier fulfill

\begin{equation}
  1- \mu \leq  F_i \leq 1- \epsilon \hspace{0.3cm} \forall i, 
\end{equation}

for some $0<  \epsilon  <\mu < 1$.

Then, the swap-test-based verification algorithm leads to a probability $p_{\mathrm{TA}}$ of true acceptance fulfilling the condition

\begin{equation}
   \Bigg(\frac{2 -\mu}{2} \Bigg)^M \leq   p_{\mathrm{TA}} \leq \Bigg(\frac{2 -\epsilon}{2} \Bigg)^M,
   \label{upper-bound_zero}
\end{equation}

having that $M$  also constrains the probability  $p_{\mathrm{TR}}$ of true rejection, as

\begin{equation}
    p_{\mathrm{TR}} \geq  1 -\Bigg(\frac{1 +\frac{d+1}{D}}{2} \Bigg)^M,
    \label{fast_pace}
\end{equation}

where $D$ is the dimension of the underlying Hilbert space, and $d$ is the dimension of the largest subspace spanned by the set of challenges queried by an adversary.

For the gswap-test-based verification algorithm, instead, we find

\begin{equation}
   \frac{1}{M+1} + \frac{M}{M+1}(1 - \mu) \leq  p_{\mathrm{TA}} \leq  \frac{1}{M+1} + \frac{M}{M+1}(1 - \epsilon).
\end{equation}
That is, in this case, $p_{\mathrm{TA}}$ is not upper-bounded by a quantity that approaches zero exponentially in $M$, as in Equation \eqref{upper-bound_zero} and, moreover, we observe an informative lower bound for it. However, for $p_{\mathrm{TR}}$, we find 

\begin{equation}
    p_{\mathrm{TR}} \geq 1 -  \frac{1}{M+1} - \frac{M}{M+1}\frac{d+1}{D}.
\end{equation}

Hence, we only find it to be lower-bounded by a quantity that   approaches $1$ at a slower pace than the one shown in Equation \eqref{fast_pace}.

Finally, notice that if larger values of $N$ are set, in order to enhance soundness, completeness is exponentially affected. That is, $p_{\mathrm{TA}}$ approaches zero exponentially fast in $N$ for both types of tests considered.

\item Secondly, the work in \cite{t-design} makes different relevant contributions to the field. On the one hand, the cited authors note that \cite{Mina_unitary_qpuf} lacks a uniqueness proof for their QPUF theoretical construction. Nevertheless, they show that uniqueness is guaranteed by the Haar-randomness hypothesis. Notably, they additionally prove that other generation strategies can also deliver uniqueness. Furthermore this article comments on the inconvenience of the Haar-randomness requirement, and proposes an alternative relying on the well-known properties of $t$-designs, delivering the first  application of this concept for a general value of $t$. The proper functioning of this alternative, however, comes at the price of restricting the number of queries by the adversary to $t$, instead of it being any polynomial amount of certain security parameters. As a final remark, this work does not omit a discussion on the effect of quantum noise, but they restrict it to the case of unitary noise channels. In such case the proposed scheme remains functional.

\item As the third and last proposal building on \cite{Mina_unitary_qpuf}, the authors of \cite{MB_QPUF} develop different schemes that further explore the potential of QPUFs, actively exploiting the Haar-randomness assumption. The contributions of this work are two-fold: on the one hand, the introduction of the Ideal QPUF model achieves the strongest kind of  unforgeability against polynomial adversaries, i.e., quantum existential unforgeability, by harnessing the randomness provided by quantum measurements. Moreover, for this proposal, multiple swap tests are no longer required for the  verification algorithm. The new acceptance procedure, instead, benefits from a one-shot scheme that owns similar desirable properties as those of the gswap verification.  Nonetheless, the proposed implementations of such Ideal PUF suffer from serious practical drawbacks including exponential circuit depth and the requirement of inverting an unknown unitary evolution. On the other hand, the derived Measurement-Based PUF (MB-QPUF) takes advantage of the properties of maximally entangled states of arbitrary dimension in order to achieve selective unforgeability, while avoiding the costly resource of Haar randomness for the challenge selection. In such case,  desirably, the two entangled system parties can be kept close to each other, avoiding the need of sustaining entanglement over large distances. 

As a final observation before concluding this section, we aim to stress how, as pointed out by the last cited authors, QPUF models require further investigation when considering noisy environments. It remains an open question whether there exist error correction procedures able to maintain the desirable security features  of the introduced QPUF schemes while rendering them robust under realistic conditions, given the current and near-term hardware limitations.
\end{enumerate}

\begin{table*}[t]
    \centering
    \caption{Comparison between  proposals of novel schemes building on \cite{Mina_unitary_qpuf}.}
    
    \begin{tabular}{c|c|c|c}
        Article & Proposal name  & Introduced improvement/s w.r.t. \cite{Mina_unitary_qpuf} & Drawback/s \\ \specialrule{0.15em}{0em}{0em}

        M. Doosti et al. \cite{client_server} (a) & HRV & gswap \& Remote authentication & Haar randomness still  \\ & & &  required  for challenge \\ & & & selection \\ \hline 
        M. Doosti et al. \cite{client_server} (b) & LRV & Low-resources verifier \&  & Haar randomness still  \\ &  & Remote autehentication & required for challenge ´ \\ & & & selection \\ \hline 
        N. Kumar et al. \cite{t-design} & QPUF from  & Haar randomness not required \&  & Vulnerable against \\ & unitary $t$-designs & Robust against unitary noise & polynomial adversaries \\ \hline 

       S. Ghosh et al. \cite{MB_QPUF} (a) & Ideal QPUF & Existentially unforgeable \& & Haar randomness still \\ &  & Haar randomness not required  & required for  QPUF generation  \& \\ & & for challenge selection \&  & Exponential circuit complexity/ \\ & & One-shot verification & unknown unitary inversion \\ \hline 

       S. Ghosh et al. \cite{MB_QPUF} (b) &  MB-QPUF & Haar randomness not required  &  Haar randomness still    \\ & & for challenge selection  & required for QPUF generation \\ 
    
    \end{tabular}
    
    \label{tab:improvements/drawbacks}
\end{table*}

\subsection{Hybrid PUFs}
In an effort to improve the practicality of weak PUFs, the authors in \cite{mina_hybrid_1} propose the use of a quantum lock: a mechanism that combines classical and weak classical PUFs with a quantum encoding based on non-orthogonal states. The quantum lock enhances the security of weak PUFs by ensuring that even if an adversary has fully characterized the underlying classical PUF, the overall construction—referred to as a Hybrid PUF (HPUF)—remains secure. This approach also enables challenge reusability, a feature typically absent in PUF designs. On the other hand, both  the \textit{offline} and \textit{online  protocols} presented in \cite{mina_hybrid_2} constitute two other HPUF models that, instead of relying on non-orthogonal state preparation, exploit the properties of maximally entangled states. 

The above proposals represent a step towards requiring more realistic resources (see Table \ref{tab:comparison} for a  key-feature comparison among all different PUFs existent today). For instance, they neither require quantum memories nor Haar randomness. Nevertheless, neither of the two articles discusses the effect of quantum noise or the threat posed by phishing-attack schemes, which can presumably compromise their security with the following strategy: the attacker impersonates the client first, in order to receive a challenge, and the server later, in order to obtain a valid response from the client and redirect it to the server.
\label{review_qpufs}

\begin{table*}[h]
    \centering
    \caption{Comparison between different features of the different kinds of PUFs existent today.}
    
    \begin{tabular}{c|c|c|c|c}
         & CPUFs  & QR-PUFs & QPUFs & HPUFs\\ \specialrule{0.15em}{0em}{0em}

        Trusted certifier & Yes & Yes  & No & Yes \\ 
        \hline 
        Security type  & Heuristic & Heuristic &  Mathematically & Heuristic \\  & & & grounded &  \\ \hline 
        Commercially  & Yes & No (to the best  & No (to the best of  & No (to the best  \\ available today & & of our knowledge) &  of our knowledge) & of our knowledge) \\  \hline 

       Experimentally  & Yes & Yes & No (to the best  & Yes \\ available today & & & of our knowledge) &    

    \end{tabular}
    
    \label{tab:comparison}
\end{table*}

\subsection{Information-theoretic analysis of QPUFs}

The security analysis of QPUFs is predominantly grounded in information-theoretic frameworks. Unlike computational security, which relies on hardness assumptions, information-theoretic approaches provide unconditional guarantees on critical properties such as unpredictability, unclonability, and entropy. Such an analysis establishes performance and security limits that are independent of implementation and adversarial capabilities, including those of quantum-capable adversaries.

Our article-selection criteria include the work in \cite{info_theore_puf}, which introduces an information-theoretic framework for QPUFs based on bipartite Discrete Memoryless Multiple Sources (DMMS), an abstraction derived from biometric source models \cite{ignatenko2012biometric}. Within this framework, QPUFs are modeled as stochastic sources, where a challenge is mapped to a response via a probabilistic transformation influenced by quantum or device-specific noise, and the challenge (input), response (output), and internal randomness are modeled as random variables. Unlike spatially distributed DMMS models, here the observations occur sequentially (in time) under varying physical conditions. This abstraction, which treats the QPUF as a black-box oracle, decouples the analysis from specific physical implementations, thereby enabling general, device-agnostic evaluations of uncertainty, noise, and information flow. It accommodates both classical and quantum challenge schemes while assuming quantum responses, encompassing a broad class of QPUF protocols \cite{nilesh2024quantum}. This analysis relies on two key bounds: the achievability bound, which ensures the existence of coding or challenge–response schemes that attain a target performance with vanishing error probability, and the converse bound, which defines the theoretical maximum that no scheme can exceed. Together, these bounds tightly determine the capacity and ultimate limits of QPUF-based cryptographic systems.

Within this framework, authentication performance is characterized via standard security metrics such as the False Acceptance Rate (FAR), representing the probability of an adversary being falsely accepted, and the False Rejection Rate (FRR), representing the likelihood of a legitimate prover being rejected. It has been shown that QPUF-based systems can be designed to achieve asymptotically vanishing FRR and exponentially decaying FAR, ensuring robust security guarantees even in the presence of quantum-capable adversaries \cite{nilesh2025authentication}.

Information-theoretic methods also underpin secure key generation and data storage using QPUFs \cite{patent1, patent2}. The core insight is that QPUFs' intrinsic randomness supplies high entropy that can be harvested into secret keys with provable secrecy. Finally, the upper bound on the number of QPUFs that can be unambiguously identified within a network while maintaining reliable authentication and secure key generation is derived in \cite{nilesh2026sub}. These high-entropy outputs can be processed (e.g., via privacy amplification) to yield secret keys. Information-theoretic analysis precisely quantifies how many secret bits can be extracted and how much information an adversary can learn. The secret key rate was shown to be related to the FAR, thereby establishing that a higher secret key rate extracted from a QPUF corresponds to a tighter bound on the adversarial success probability \cite{nilesh2025authentication}. In \cite{info_theore_puf, nilesh2025quantum} secret key capacities extractable from QPUF outputs under privacy leakage have been bounded via mutual-information expressions. Extensions to secure data storage and message identification (detecting if a specific message is present instead of full decoding) via QPUFs were further characterized in \cite{nilesh2025itw, nilesh2025secure, nilesh2025globe}, wherein identification capacity was shown to scale double-exponentially with the QPUF output length under idealized conditions. Importantly, these entropy-based bounds hold regardless of adversarial computational power. An information-theoretic argument guarantees that even a quantum-enabled adversary cannot reduce the QPUF’s inherent entropy below its limit. In turn, the QPUF’s intrinsic randomness is tied directly to provable secrecy: for example, any key derived from the QPUF can be made statistically independent of public helper data. These security guarantees rely on the concept of privacy leakage, which requires that public helper data reveal negligible information about the QPUF’s output. This requirement is especially critical in quantum settings, as compromise of QPUF output could lead to irreversible identity theft.

All of these results rely on idealizations, i.i.d. noise, perfect challenge-response correlations, and unbounded blocklength, that abstract away practical limitations such as hardware imperfections, correlated errors, and finite‐length effects. The necessity of bridging the gap between theoretical information‐theoretic security proofs and realistic implementations has been emphasized in recent discussions \cite{amiri2025quantum, amiri2025quantum2, nilesh2025fnwf}, and remains an important direction for future work. Despite these caveats, the information-theoretic framework provides a rigorous, implementation-independent foundation for the analysis of QPUF‐based cryptographic protocols, ensuring robustness even against adversaries endowed with quantum capabilities.

\section{Outlook on future research}
\label{future}

We conjecture that novel approaches may yet emerge within the QPUF
literature. In analogy with the historical development of classical communication, one may
envisage the possibility of a transformative idea in which quantum noise is harnessed as a
resource rather than regarded exclusively as a hindrance \cite{WIRETAP, diff_power_analysis}. Since the field remains in an early
stage of development and has not yet yielded fully satisfactory solutions to the problems it
seeks to address, the exploration of new perspectives and methodologies appears both
necessary and timely.

Future work is needed to define QPUF norms that can rigorously evaluate and certify QPUF devices for real-world deployment. Although significant progress has been achieved in the design and demonstration of QPUFs, the field currently lacks standardized frameworks for characterization, benchmarking, and interoperability across architectures such as photonic, superconducting, and ion-trap implementations. This absence impedes reproducibility, cross-platform comparison, and integration of QPUFs into broader quantum communication and cryptographic infrastructures. Existing evaluation metrics such as uniqueness, reliability, and unclonability remain inconsistently defined across experimental platforms, and no unified test protocols or certification criteria have been established. Future standardization efforts must therefore develop architecture-agnostic metrics, interoperable quantum–classical interfaces, rigorous verification methodologies, and seamless authentication across multi-vendor networks. Such standards will be essential for transitioning QPUFs from isolated laboratory prototypes to trusted, scalable components in quantum-secure hardware ecosystems. Furthermore, the high cost of quantum components currently restricts QPUFs to high-security domains like critical infrastructure and defense. Realizing QPUFs at scale will require coordinated advances across multiple disciplines: materials science and quantum engineering to enhance coherence, reduce noise, and integrate photonic or cryogenic components into compact, SWaP-C compliant packages.

\section{Conclusions}
\label{conclusions}

In this review article, we  explore the concept of Quantum Physical Unclonable Functions (QPUFs) from both theoretical and practical perspectives. The initial theoretical framework serves as a crucial starting point for formalizing the essential criteria a QPUF must satisfy. However, while conceptually well defined, this model faces substantial experimental limitations that currently hinder its practical deployment. Among the most pressing challenges are the handling of quantum noise and the development of an effective and reliable QPUF generator. However, a range of studies have emerged that build upon this foundational framework, either by strengthening its security properties or by relaxing some of its more demanding practical constraints. These contributions offer valuable insights into how theoretical models might evolve towards implementable forms.

Our review also considers Quantum Readout PUFs (QR-PUFs), which represent a more practical and experimentally demonstrated class of devices. Although promising, these implementations rely on certain strong assumptions, and their security may be compromised when those assumptions do not hold. We also note that many models identified as QR-PUFs are frequently labeled as QPUFs in the literature. This blending of definitions reveals a notable polysemy, highlighting the need for clearer terminology and classification within the field.

Furthermore, we examine a novel solution found in recent contributions to the field: Hybrid PUFs (HPUFs). These aim to extend the capabilities of classical PUFs by integrating quantum features. These models show potential for supporting authentication protocols and represent a step towards practical applications grounded in classical-quantum hybrid systems.

Finally, we review the main works in the state of the art of studying QPUFs via information-theoretic analyses, which typically focus on both achievability and converse bounds, and exploit the interconnection between secret-key generation and authentication. We further acknowledge the idealizations that are typically present in approaches of this nature, and comment on how these affect the actual implications of the outcomes implied by such  studies.

\sloppy

Before concluding this article, we want to further notice how, in this work, we have discussed how some quantum-based PUFs—such as QR-PUFs—can still be regarded as partly classical. Interestingly, the converse also holds: certain commercial implementations of classical PUFs, such as those by Crypto Quantique \cite{cryptoquantique2024qdid} and PUFsecurity \cite{pufsecurity2020neopuf}, rely on fabrication-level variations originating from quantum phenomena, including tunneling in Complementary Metal–Oxide–Semiconductor (CMOS) or Metal–Oxide–Semiconductor Field-Effect Transistor (MOSFET) devices. Moreover, across all forms of PUFs, ensuring True Random-Number generation appears essential for security and unpredictability, as highlighted in recent industrial developments integrating quantum-based randomness \cite{xiphera2024qdid, ictk2021idq}. Quantum technologies can contribute significantly to this critical cryptographic primitive \cite{TRN1, TRN2}.

\section*{Acknowledgements}

The authors thank Hadi Aghaee for valuable discussions.
The authors further gratefully acknowledge financial support from the Federal Ministry of Research, Technology and Space of Germany (BMFTR) within the program Souverän. Digital. Vernetzt. This work was supported by the joint project 6G-life, project identification numbers 16KISK002 and 16KISK263. Furthermore, VG and JN acknowledge additional support from BMFTR through projects 16KISQ039, 16KISQ077, and 16KISQ168, as well as from the German Research Foundation (DFG) under project NO 1129/2-1. CD and PJF received further funding from BMFTR projects 16KISQ169, 16KIS2196, 16KISQ038, 16KISR038, 16KISQ170, and 16KIS2234. HB and KN were supported by BMFTR projects 16KISQ037K, 16KISQ077, 16KISQ093, 16KISR026, 16KIS2195, and 16KIS1598K. TS, GP, and MB acknowledge support from the BMBF projects QPIS (No. 16KISQ032K) and DINOQUANT (No. 13N14921), as well as from the European Commission through the ERC Starting Grant project QUREP (No. 851810).

\section*{Appendix: Article-selection table for the review}

Table \ref{tab:quantum_puf} shows the list of articles considered in the review, including: title, authors, year, citation count, reference count and similarity score. 

\begin{landscape}
\begin{table}[h]
\vspace{-8mm}
    \centering
    \small
    \begin{tabular}{l l c c c c}
        \toprule
        \textbf{Title} & \textbf{Authors} & \textbf{Year} & \textbf{Citations} & \textbf{References} & \textbf{Similarity} \\
        \midrule

            \colorrowd
\cite{mina_hybrid_2} Hybrid Authentication Protocols for Advanced Quantum Networks & Suchetana Goswami et al. & 2025 & 0 & 76 & 1.00 \\
        \cite{li2025quantumPUF} QPUF Based on Multidimensional Fingerprint Features of Single Photon Emitters.. & Qian Li et al. & 2025 & 3 & 49 & 0.00 \\

        \cite{Huang2024} Near-Infrared Circularly Polarized Luminescent Physical Unclonable Functions & Jiang Huang et al. & 2024 & 13 & 14 & 0.00 \\

        \cite{bathalapalli2024qpuf20exploringquantum} QPUF 2.0: Exploring Quantum Physical Unclonable Functions ... & Venkata K. V. V. B. et al. & 2024 & 0 & 55 & 0.07 \\

        \cite{MARYMATHEWS2024103787} QS-Auth: A Quantum-secure mutual authentication... & Mahima Mary Mathews et al. & 2024 & 1 & 81 & 0.05 \\
        
        \cite{10373567} Soteria: A Quantum-Based Device Attestation Technique... & Mansoor Ali Khan et al. & 2024 & 10 & 46 & 0.05 \\
                    \colorrowd

        \cite{MB_QPUF} Existential unforgeability in quantum authentication... & Soham Ghosh et al. & 2024 & 4 & 41 & 1.00 \\
                            \colorrow

        \cite{info_theore_puf} Information Theoretic Analysis of a Quantum PUF & Kumar Nilesh et al. & 2024 & 1 & 28 & 0.29 \\
                                    \colorrow

        \cite{weak_QPUF} QPUF: Quantum Physical Unclonable Functions for Security-by-Design... & Venkata K. V. V. B. et al. & 2023 & 3 & 23 & 0.13 \\
        
        \cite{nocentini2023physicalrealizationhyperunclonable} Physical Realization... & Sara Nocentini et al. & 2023 & 1 & 41 & 0.05 \\
                            \colorrow

        \cite{q_lock} Quantum Logic Locking for Security & Rasit Onur Topaloglu & 2023 & 6 & 16 & 0.13 \\

        \cite{10365761} Quantum Crosstalk as a Physically Unclonable Characteristic... & Christopher Z. Chwa et al. & 2023 & 2 & 18 & 0.06 \\
                            \colorrow

        \cite{hardware_with_errors_qrpuf_2} Trustworthy Quantum Computation through...
         & Kaitlin N. Smith et al. & 2023 & 0 & 28 & 0.18 \\

                                     \colorrow

        \cite{QPUFs_comparison} Comparison of Quantum PUF models & Vladlen Galetsky et al. & 2022 & 9 & 29 & 0.13 \\

        \cite{cryptoeprint:2020/237} On Security Notions for Encryption in a Quantum World & C. Chevalier et al. & 2022 & 40 & 38 & 0.05\\
        
        \cite{Katumo2022} Dual-color dynamic anti-counterfeiting labels with persistent... & Ngei Katumo et al. & 2022 & 33 & 49 & 0.02\\

        \cite{cryptoeprint:2022/699} On the Quantum Security of OCB & Varun Maram et al. & 2022 & 7 & 45 & 0.02 \\
                                    \colorrow

        \cite{hardware_with_errors_qrpuf1} Quantum PUF for Security and Trust... & Koustubh Phalak et al. & 2021 & 73 & 18 & 0.28 \\
        
        \cite{Mina_Pseudorandomness} On the connection between quantum pseudorandomness... & Mina Doosti et al. & 2021 & 5 & 48 & 0.08 \\
        
                                    \colorrow
\cite{framework_unforgeability} A Unified Framework For Quantum Unforgeability & Mina Doosti et al. & 2021 & 12 & 34 & 0.15 \\
                                    \colorrow

        \cite{mina_hybrid_1} Quantum Lock: A Provable Quantum Communication Advantage & Kaushik Chakraborty et al. & 2021 & 8 & 70 & 0.34 \\
                                    \colorrow

        \cite{t-design} Efficient Construction of Quantum Physical Unclonable Functions... & N. Kumar et al. & 2021 & 14 & 41 & 0.10 \\
                                    \colorrow

        \cite{learning_qrpuf} Learning classical readout quantum PUFs based on single-qubit gates & Anna Pappa et al. & 2021 & 8 & 21 & 0.24 \\
                                    \colorrow

        \cite{cpuf_for_authentication} Remote quantum-safe authentication of entities... & G. Nikolopoulos & 2021 & 6 & 32 & 0.19 \\
                                    \colorrow

        \cite{client_server} Client-server Identification Protocols with Quantum PUF & Mina Doosti et al. & 2020 & 24 & 58 & 0.16 \\
        
        \cite{Gu2020} Gap-enhanced Raman tags for physically unclonable anticounterfeiting labels & Yuqing Gu et al. & 2020 & 227 & 64 & 0.02 \\
                                    \colorrow

        \cite{964b3cf542769a9fdb6544165848f47a203b6382} Security Analysis of Identification Protocols... & Frederick Hetherton & 2020 & 0 & 30 & 0.20 \\
        
        \cite{e51e0cc3aba2c17c5e543a27b91fef0793913e04} Analysis of crosstalk in NISQ devices... & Abdullah Ash-Saki et al. & 2020 & 64 & 14 & 0.00 \\
                    \colorrowd

        \cite{Mina_unitary_qpuf} Quantum Physical Unclonable Functions: Possibilities and Impossibilities & Myrto Arapinis et al. & 2019 & 55 & 55 & 1.00 \\
        
        \cite{Zhang2019} PbS Quantum Dots Based on PUFs for Ultra High-Density Key Generation & Yuejun Zhang et al. & 2019 & 5 & 29 & 0.00 \\
                                    \colorrow

        \cite{framework_readout} Theoretical framework for physical unclonable... & Giulio Gianfelici et al. & 2019 & 24 & 41 & 0.20 \\
        
        \cite{Nikolopoulos_2019} Optical scheme for cryptographic commitments with physical unclonable keys & Georgios M. Nikolopoulos & 2019 & 5 & 33 & 0.06 \\
                            \colorrow
        
        \cite{Fladung_2019} Intercept-Resend Emulation Attacks Against a Continuous-Variable...& L. Fladung et al. & 2019 & 14 & 35 & 0.09 \\
        
        \cite{62a58f3eae89c1637ded1dd2236733b695c7ce64} (Tightly) QCCA-Secure Key-Encapsulation... & Keita Xagawa et al. & 2019 & 1 & 24 & 0.04 \\
        
        \cite{fc16e68ed2840255ecad5c758d77b65e90967023} A quantum related-key attack based... & H. Xie, L. Yang & 2018 & 16 & 52 & 0.00 \\
        
        \cite{Nikolopoulos2017} Continuous-variable quantum authentication of... & G. Nikolopoulos & 2018 & 47 & 37 & 0.03 \\
        
        \cite{QRPUF_Skoric_4} Asymmetric cryptography with physical unclonable keys & R. Uppu et al. & 2018 & 34 & 38 & 0.05 \\

        \cite{8010947} A Retrospective and a Look Forward: Fifteen Years of... & Chip-Hong Chang et al. & 2017 & 177 & 132 & 0.05 \\
                            \colorrow

        \cite{QRPUF_Skoric_3} Authenticated communication from quantum readout of PUFs & B. Škorić et al. & 2017 & 22 & 12 & 0.25 \\
                                    \colorrow

        \cite{a7d2ee8303c4e992579a136494c3fa8d0f450373} Towards a Unified Security Model for... & F. Armknecht et al. & 2016 & 63 & 32 & 0.16 \\
                                    \colorrow

        \cite{clone_attacks_qrpuf} Quantum cloning attacks against PUF-based... & Y. Yao et al. & 2016 & 22 & 34 & 0.18 \\
                                    \colorrow

        \cite{security_challenge} Security analysis of Quantum-Readout PUFs in the... & B. Škorić & 2016 & 14 & 19 & 0.37 \\
                                    \colorrow

        \cite{652092ac481b283c4f05134f4f65e2f9d159b44e} Using Quantum Confinement to... & J. Roberts et al. & 2015 & 54 & 38 & 0.13 \\
                                    \colorrow

        \cite{learning_cpufs} PUF modeling attacks: An introduction and overview & U. Rührmair, J. Sölter & 2014 & 112 & 51 & 0.18 \\
                                    \colorrow

        \cite{quadrature_attacks} Security of Quantum-Readout PUFs against quadrature-based... & B. Škorić et al. & 2013 & 32 & 26 & 0.23 \\
                                    \colorrow

        \cite{experimental_qrpuf} Quantum-secure authentication of a physical... & S. A. Goorden et al. & 2013 & 222 & 23 & 0.17\\
                                    \colorrow

        \cite{bit_commitment} On the practical use of physical unclonable functions in... & U. Rührmair, M. van Dijk & 2013 & 52 & 37 & 0.22 \\
                            \colorrow

        \cite{sharp_lower_bounds} Sharp lower bounds on the extractable randomness from non-uniform sources & B. Škorić et al. & 2011 & 15 & 23 & 0.17 \\
                    \colorrowd

        \cite{readout_qpuf} Quantum Readout of Physical Unclonable Functions & B. Škorić & 2010 & 74 & 31 & 1.00 \\
        \bottomrule
    \end{tabular}
    \caption{QPUF papers are sorted by publication date (as of May 12th, 2025). Baseline articles are marked in yellow; accepted ones (with threshold $l = 0.1$) appear in light gray.}
    \label{tab:quantum_puf}
\end{table}

\end{landscape}

\medskip

%

\end{document}